\documentclass[a4paper,UKenglish,cleveref, autoref, thm-restate]{lipics-v2021}
\pdfoutput=1 %uncomment to ensure pdflatex processing (mandatatory e.g. to submit to arXiv)
\hideLIPIcs  %uncomment to remove references to LIPIcs series (logo, DOI, ...), e.g. when preparing a pre-final version to be uploaded to arXiv or another public repository

\usepackage{algorithm}
\usepackage{algpseudocode}
\usepackage{makecell}
\usepackage{multirow}

\title{Generating Realistic Individual Activity Schedules via Activity Location Allocation Based on Simulated Travel Times} %TODO Please add

\titlerunning{Generating Realistic Individual Activity Schedules} %TODO optional, please use if title is longer than one line

\author{Tatsuya Mitomi}{AI for Collective Intelligence Hub, University of Glasgow, Glasgow, UK \and Fujitsu Limited, Kawasaki, Japan}{tatsuya.mitomi@glasgow.ac.uk}{https://orcid.org/0009-0007-1195-866X}{}%TODO mandatory, please use full name; only 1 author per \author macro; first two parameters are mandatory, other parameters can be empty. Please provide at least the name of the affiliation and the country. The full address is optional. Use additional curly braces to indicate the correct name splitting when the last name consists of multiple name parts.

\author{Yahya Gamal}{AI for Collective Intelligence Hub, University of Glasgow, Glasgow, UK} {Yahya.Gamalaldin@glasgow.ac.uk}{https://orcid.org/0000-0003-2370-6172}{}

\author{Esra Suel}{Urban Analytics, Department of Geography, University of Zurich, Zurich, Switzerland}{esra.suel@geo.uzh.ch}{https://orcid.org/0000-0001-9246-3966}{}

\author{Gary Polhill}{Information and Computational Sciences Department, The James Hutton Institute, Aberdeen, UK}{Gary.Polhill@hutton.ac.uk}{https://orcid.org/0000-0002-8596-0590}{}

\author{Daniel Konioukhov}{AI for Collective Intelligence Hub, University of Glasgow, Glasgow, UK}{3174941K@student.gla.ac.uk}{https://orcid.org/0009-0007-8198-9875}{}

\author{Alison Heppenstall}{AI for Collective Intelligence Hub, University of Glasgow, Glasgow, UK} {Alison.Heppenstall@glasgow.ac.uk}{https://orcid.org/0000-0002-0663-3437}{}

\authorrunning{T. Mitomi et al.} %TODO mandatory. First: Use abbreviated first/middle names. Second (only in severe cases): Use first author plus 'et al.'
\Copyright{Tatsuya Mitomi, Yahya Gamal, Esra Suel, Gary Polhill, Daniel Konioukhov, and Alison Heppenstall} %TODO mandatory, please use full first names. LIPIcs license is "CC-BY";  http://creativecommons.org/licenses/by/3.0/

%TODO mandatory: Please choose ACM 2012 classifications from https://dl.acm.org/ccs/ccs_flat.cfm 

 %TODO mandatory; please add comma-separated list of keywords

\nolinenumbers %uncomment to disable line numbering

\begin{document}

\maketitle

\begingroup
\renewcommand{\thefootnote}{}
\footnotetext{This is the author version of a short paper accepted for presentation in the poster session at the 17th Conference on Spatial Information Theory (COSIT 2026).}
\endgroup

%TODO mandatory: add short abstract of the document
\begin{abstract}
Individual level daily activity schedules are essential for a wide range of applications, including infectious disease control, urban transportation planning, and policy design. 
In practice, such schedules are typically generated by combining population data with travel survey data.
These data sources are used because they are often publicly available, whereas observed individual activity schedules are difficult to obtain due to privacy concerns.
However, because of the complexity of mobility modelling, it is difficult to generate realistic activity schedules that also preserve travel times consistent with those reported in travel surveys.
To address this issue, we propose a framework for generating activity schedules that iteratively applies a dynamic programming method to allocate activity locations based on simulated travel times.
Numerical experiments with dummy data show that the proposed method reduces the discrepancy between simulated travel times and those reported in travel surveys by 52.2\% relative to the first iteration through iterative refinement.
\end{abstract}

\section{Introduction}
\label{sec:typesetting-summary}
Datasets of individual level daily activity schedules, together with sociodemographic attributes such as gender and age, are critical for a wide range of applications, including infectious disease control, urban transport planning, and policy formulation~\cite{WANG2022103939}.
However, such datasets are difficult to obtain in practice due to privacy concerns and limited observability. These data are typically generated using other observable data sources.

\begin{figure}[t] 
  \vspace{-10mm} 
  \centering
  \includegraphics[width=\linewidth]{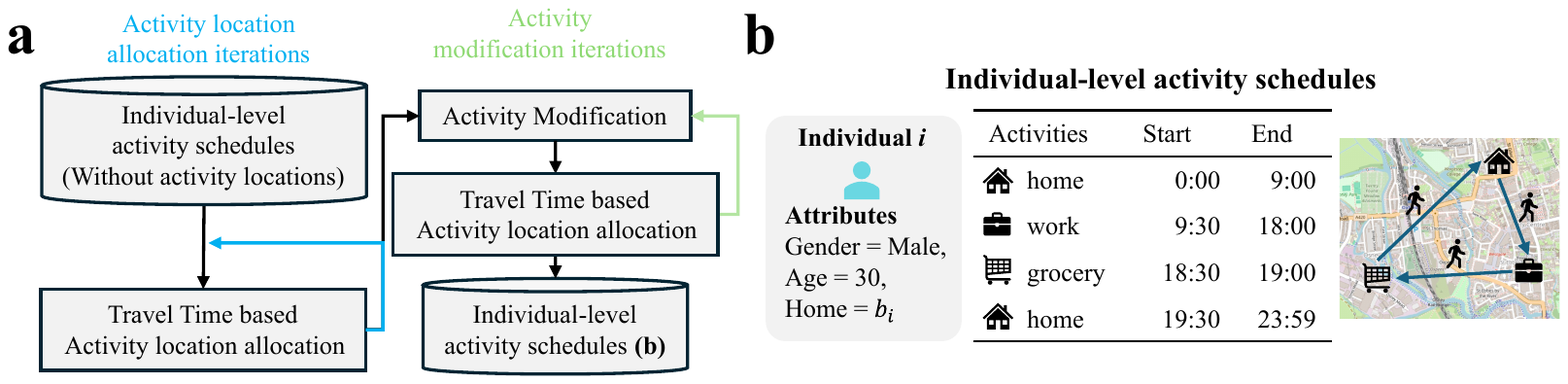} 
  \vspace{-65mm}
  \caption{Overview of the proposed framework. (a) Overall workflow of the method. The location allocation iteration is the process proposed in this study, whereas the activity modification iteration extends an existing activity modification framework~\cite{GeoAI}.
(b) Example individual-level activity schedules.}
  \vspace{-3mm}
  \label{fig:Overall_method}
\end{figure}

To this end, several studies have proposed methods for generating individual level activity schedules (Figure~\ref{fig:Overall_method}b) using population data and travel survey data~\cite{HORL2021103291, mahfouz2025reproducible}.
These generation methods often incorporate observed static travel times between location pairs rather than mobility simulation-based travel times.
However, reliance on static travel times may overlook uncertainty in activity locations arising from the complex and dynamic nature of traffic conditions. 
Because activity locations determine an individual’s travel path, they also influence congestion patterns and travel times -- these are often unaccounted for by existing methods. 
Therefore, relying solely on static travel time data may lead to unrealistic activity locations and, consequently, implausible travel times that do not match those observed in the travel survey.
Since activity locations are endogenous to the generation of daily activities, they can be modified by using mobility simulation-based travel time in order to generate more realistic individual level activity schedules.

To address this issue, we propose a framework for generating activity schedules that combines iterative refinement with a dynamic programming method for allocating activity locations based on simulated travel times.
Additionally, we investigate how this framework can be integrated into an existing activity modification framework~\cite{GeoAI} to further improve the matching of travel times between the simulated and the travel survey data.
Numerical experiments with dummy data demonstrate that the proposed framework can generate activity schedules and, through iterative refinement, reduce the discrepancy between simulated and travel survey data by 52.2\%.
These results demonstrate the potential of this approach to produce realistic individual level activity schedules, enabling more accurate simulations for transport planning.

\section{Method}
We use population data and travel survey data as the starting point, as these data sources are often publicly available as open data~\cite{Rice2025, nts2025, HORL2021103291}. 
As a first step, a statistical matching procedure is applied to combine the population data with the travel survey data~\cite{HORL2021103291}. This matched dataset is used as the input to the proposed method. 
In general, population data include individual attributes such as gender, age and home locations, whereas travel survey data include daily activity types, start times and durations, as well as travel times and travel modes between activities.
As a result, the combined dataset contains all the necessary information to generate activity schedules except activity locations.

The specific objective of this framework is to estimate activity locations in the combined dataset to ensure that simulated travel times between activities match those observed in the travel survey data. 
Figure~\ref{fig:Overall_method} illustrates the overall workflow of the proposed method.

\subsection{Simulated Travel Time Based Activity Location Allocation}
\begin{figure}[t] 
  \vspace{-10mm} 
  \centering
  \includegraphics[scale=0.41]{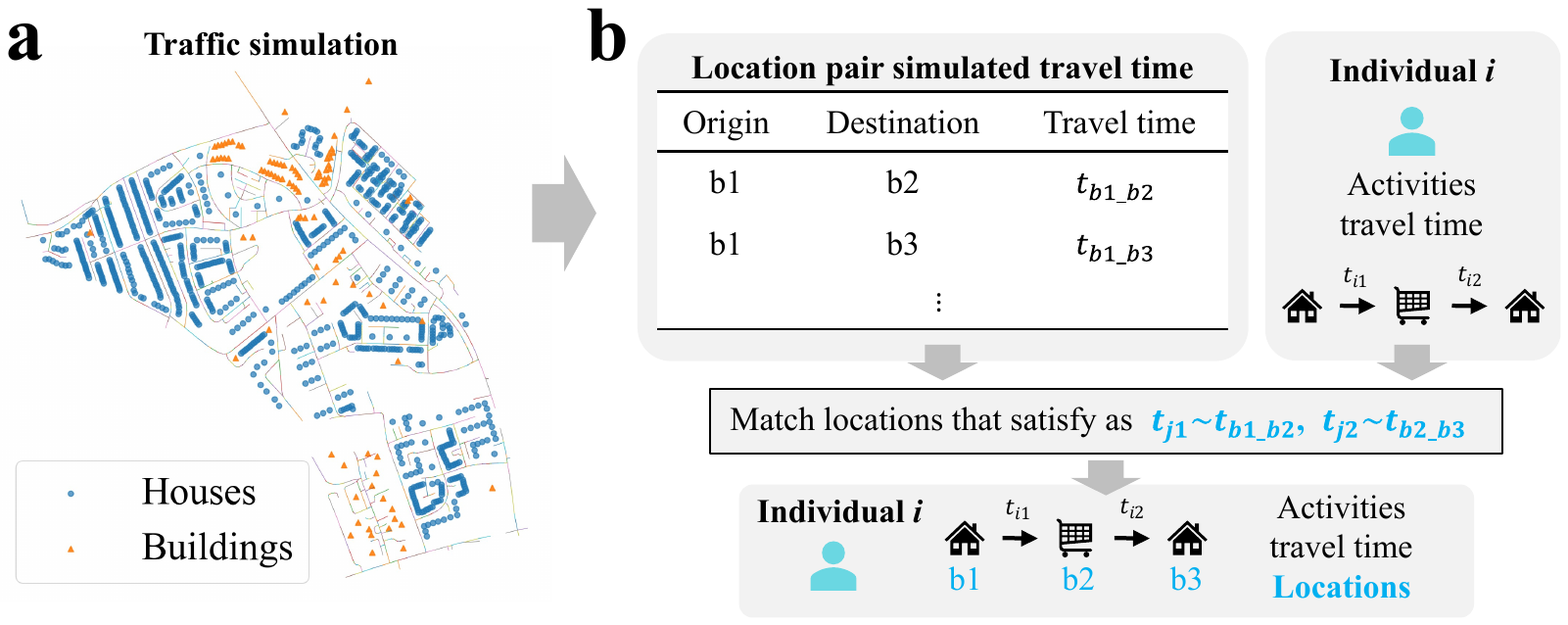} 
  \vspace{-36mm} 
  \caption{Overview of the simulated travel time based activity location allocation. (a) Example traffic simulation on a map of Aberdeen, UK, used to pilot the application of the proposed method. (b) Overview of activity location allocation using simulated travel times between location pairs.}
  \vspace{-4mm}
  \label{fig:location_matching}
\end{figure}

We introduce a dynamic programming based procedure for activity location allocation, similar to the Viterbi algorithm on a weighted graph~\cite{18626}, where dynamic programming refers to a method for solving the allocation problem through sequential multistage decomposition~\cite{P-550}.
This procedure computes a closed path of locations $(l_0,\dots,l_T)$, such that the simulated travel times along the route are as close as possible to the survey travel times ($dur_1,\dots,dur_T$) (Figure~\ref{fig:location_matching}). 
The algorithm is deterministic and it always identifies a single best route among all feasible route combinations.
To compute this route, the algorithm requires simulated travel times and individual specific travel survey information.
For the simulated travel times, let $V$ denote the set of all locations on a map, and let $W = \{ (u, v) \mid u, v \in V,\ u \neq v \}$ denote the set of paths between pairs of locations, with an associated travel time function $w : W \to \mathbb{R}$ such that $w(u, v) = w_{uv}$. 
Regarding individual specific information, the inputs are: the individual's activity schedule $a_1,\dots,a_T$, the survey travel times between these activities $dur_1,\dots,dur_T$, and the individual's home location $l_{\mathrm{home}}$. 
Algorithm ~\ref{alg:viterbi_route} presents this procedure.
In the initial step of the algorithm, simulated travel times cannot be used because the traffic simulation requires activity locations (i.e., the origins and destinations of individual trips), which are themselves generated by this algorithm.
Therefore, in the initial step, we use travel times between locations calculated based on the speed limits assigned to each road segment instead of simulated travel times.

We then apply an iterative procedure consisting of the activity location allocation and traffic simulation (Figure~\ref{fig:Overall_method}a) to address the uncertainties that arise from activity locations.  These locations determine the individuals’ travel paths, which in turn affect congestion patterns and travel times. 
By iterating between the activity location allocation and traffic simulation, the simulated travel times associated with the matched locations are repeatedly fed back into the activity location allocation process.  This ensures consistency between the travel times reported in the travel survey and those produced in the simulation.

\subsection{Sociodemographic based Activity Modification}
In addition to introducing a dynamic programming procedure for activity location allocation, we investigate how the proposed method can be integrated into an existing activity modification framework~\cite{GeoAI}. 
Specifically, this framework swaps activities between pairs of individuals with similar sociodemographic attributes to enable more flexible activity location allocation.
This allows for flexible activity location allocation while preserving the empirical activity sequences and their distributions within each sociodemographic cluster. This framework is extended and integrated into the methodology in this work.

The extended activity modification framework first follows the existing method to identify candidate pairs of individuals with similar sociodemographic attributes based on the \emph{attribute space difference}~\cite{GeoAI}.
In contrast to the existing activity modification framework, our approach does not account for household structure or the geospatial locations of candidates undergoing activity switching.
The procedure then calculates, for each candidate pair $(i,j)$, the error function defined below for both the original pair and the pair obtained by swapping their activity sequences, activity start times and survey travel times. 
The resulting errors are denoted by $e_{i,j}^{orig}$ and $e_{i,j}^{swapped}$, respectively.
Unlike the function used in the existing activity modification framework, this study defines a new error function.
The error function is defined as follows:
\begin{equation}
e_{i,j}
= \sum_{t \in T} \bigl( dur^{sim}_{i,t} - dur_{i,t} \bigr)^2
+ \sum_{t \in T} \bigl( dur^{sim}_{j,t} - dur_{j,t} \bigr)^2 .
\label{pair_error}
\end{equation}
Subsequently, for each pair of individuals satisfying the following condition, their activity sequences, activity start times and survey travel times are swapped:
\begin{equation}
e_{i,j}^{swapped} < e_{i,j}^{orig} .
\label{check}
\end{equation}
The framework also employs an iterative procedure consisting of traffic simulation, activity modification and activity location allocation, as shown in Figure~\ref{fig:Overall_method}a.

\begin{algorithm}[t]

\caption{Viterbi algorithm based activity location allocation}
\label{alg:viterbi_route}
\footnotesize
\begin{algorithmic}[1]

\Require Map travel-time function $W$ with edge travel times $w_{uv}$;
         activity schedule $a_1,\dots,a_T$;
         home location $l_{\mathrm{home}}$;
         observed travel times $dur_1,\dots,dur_T$

\State Initialize $DP[t][v] \gets \infty$ for all $t, v$
\Comment{$DP[t][v]$: minimum cost to reach node $v$ at step $t$}
\State $DP[0][l_{\mathrm{home}}] \gets 0$

\State Initialize $Back[t][v]$ for all $t, v$
\Comment{$Back[t][v]$: previous node before arriving at $v$ at step $t$}
\State Initialize $Time[t][v]$ for all $t, v$
\Comment{$Time[t][v]$: edge travel time used to arrive at $v$ at step $t$}

\For{$t = 0$ to $T-1$}
    \State $d \gets dur_{t+1}$
    \ForAll{$v \in V$}
        \If{$DP[t][v] = \infty$}
            \State \textbf{continue}
        \EndIf
        \ForAll{$(v,u) \in W$}
            \State $cost \gets DP[t][v] +\ (d - w_{vu})^2$
            \If{$cost < DP[t+1][u]$}
                \State $DP[t+1][u] \gets cost$
                \State $Back[t+1][u] \gets v$
                \State $Time[t+1][u] \gets w_{vu}$
            \EndIf
        \EndFor
    \EndFor
\EndFor

\State $l_T \gets l_{\mathrm{home}}$
\State $curLoc \gets l_{\mathrm{home}}$
\For{$t = T$ down to $1$}
    \State $prevLoc \gets Back[t][curLoc]$
    \State $l_{t-1} \gets prevLoc$
    \State $curLoc \gets prevLoc$
\EndFor

\For{$t = 1$ to $T$}
    \State $dur^{sim}_t \gets Time[t][l_t]$
\EndFor

\State \Return $(l_0,\dots,l_T)$ and $(dur^{sim}_1,\dots,dur^{sim}_T)$

\end{algorithmic}
\end{algorithm}

\subsection{Traffic Simulation}
We deploy the agent-based traffic simulation framework MATraM~\cite{gamal2026matrammultiactivitytransportmobility}. 
This simulation framework models a population of individuals moving within a geospatial environment that includes a road network and building locations. 
Individuals execute daily activity schedules, travel between activity locations, and interact indirectly through congestion effects. 
Each activity is assigned to a building and trips are defined as origin-destination pairs between consecutive activities.
In this study, we use a version of MATraM in which only car agents are simulated.

% \section{Showcase of The Method}
\section{Pilot Application}
\subsection{Data and Traffic Simulation Settings}
The proposed method is verified using dummy activity schedules and a dummy population, together with a traffic map of Aberdeen, Scotland (Figure~\ref{fig:location_matching}a), downloaded from OpenStreetMap~\cite{OpenStreetMap}.
In these dummy activity schedules and population, sociodemographic attributes are sampled from empirical distributions for employment status (unemployment, full-time employment, and part-time employment), gender and age, which are based on the National Travel Survey ~\cite{nts2025} in England and Wales.
For the dummy activity schedules, we adopt the following assumptions for data generation:
(1) Activity sequences, activity types, start times, and durations are sampled from the corresponding empirical distributions.
The activity types considered are \texttt{\textbf{Work}}, \texttt{\textbf{Home}}, \texttt{\textbf{Grocery}}, and \texttt{\textbf{Entertainment}}.
(2) Travel times between activity locations are drawn from a lognormal distribution $\mathrm{Lognormal}(\mu,\sigma)$, where 
$\sigma^{2} = \ln\left(1 + \frac{0.005}{0.08^{2}}\right)$ and 
$\mu = \ln(0.2) - \frac{\sigma^{2}}{2}$. 
For simplicity, vehicles are assumed to travel along road segments at empirically specified speed limits.
The number of iterations for both activity location allocation and activity modification is set to $10$ each.
The experiment is repeated five times. 
All experiments are performed on a machine equipped with an Intel Core Ultra 7 165H CPU (3.10 GHz) and 32 GB of RAM.

\begin{figure}[t]
  \centering

  \begin{minipage}[t]{0.5\linewidth}
    \centering
    \includegraphics[width=\linewidth]{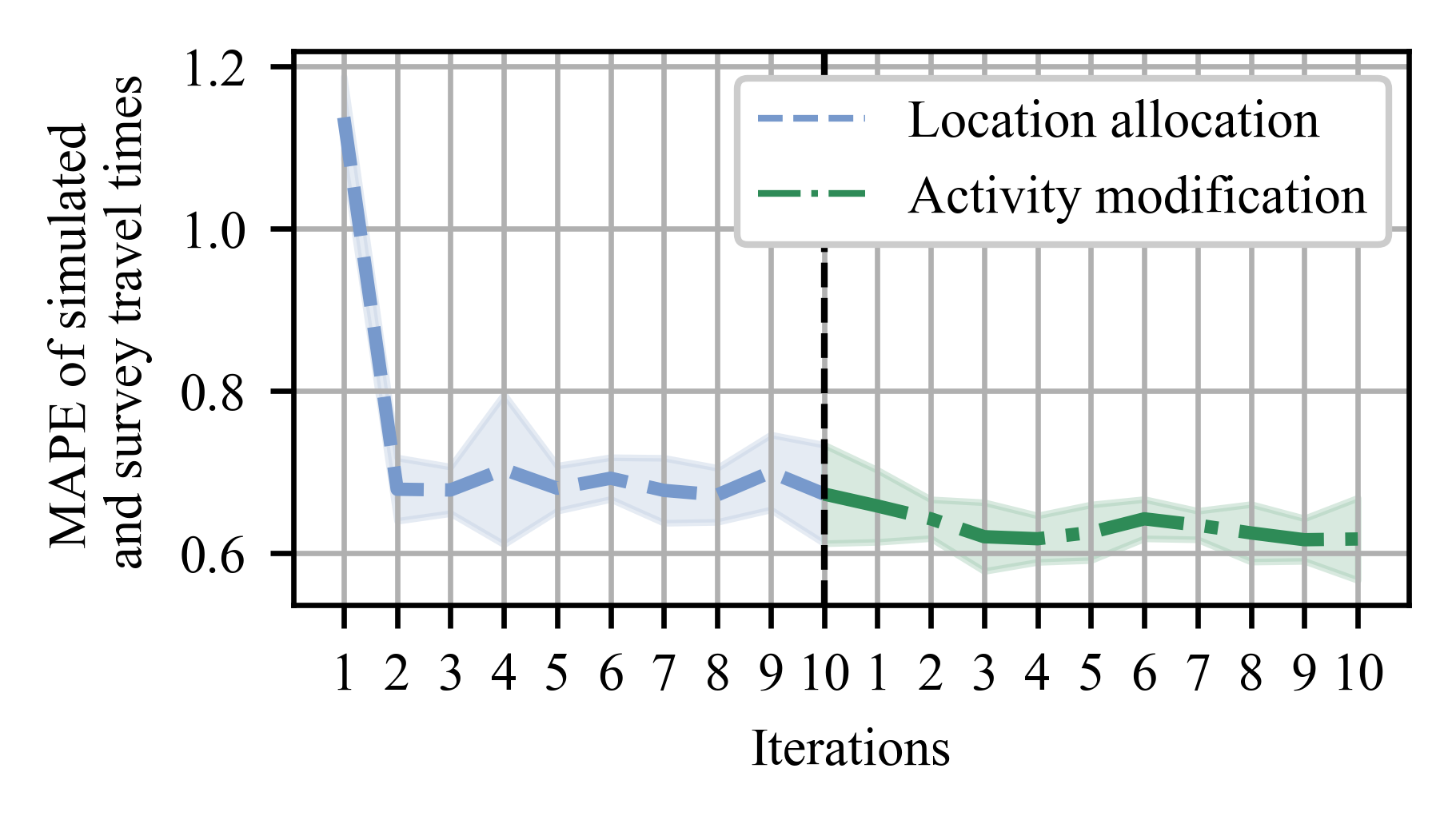}
  \end{minipage}
  \hfill
  \begin{minipage}[t]{0.49\linewidth}
    \vspace{-35mm} 
    \centering
    \begin{tabular}{ll|r}
      \hline
      Method & Iter. &  MAPE \\
      \hline
      Location allocation     & 1st   & 1.14 \\
      Location allocation     & 10th  & 0.673 \\
      Activity modification & 1st  & 0.658 \\
      Activity modification & 10th  & 0.618 \\
      \hline
    \end{tabular}
  \end{minipage}

  \caption{Change in the mean absolute percentage error (MAPE) between simulated and survey travel times, shown as the mean over five runs, with a band indicating the minimum and maximum values. The table reports the MAPE values between simulated and survey travel times at specific iterations.}
  \label{fig:mae_and_accuracy}

\end{figure}

\subsection{Results}
\subsubsection{Travel Time Matching}
Figure~\ref{fig:mae_and_accuracy} illustrates the change in mean absolute percentage error (MAPE) between simulated and survey travel times over every iteration, aggregated across all trips and individuals. 
The results for activity location allocation indicate a 46.7\% reduction in MAPE from the first to the tenth iteration. 
This shows that the proposed activity location allocation process achieves better alignment of activity locations in terms of travel time. 
In addition, the results for activity modification indicate a further 5.5\% reduction in MAPE. 
Overall, the proposed method achieves a total 52.2\% reduction in MAPE, confirming that integrating activity modification with the proposed activity location allocation method further improves the consistency between simulated and survey travel times.

\subsubsection{Sociodemographic Attributes}
Although the extended activity modification framework enables flexible activity location allocation, it can affect the relationship between sociodemographic attributes and activity sequence patterns. 
Therefore, we analyze the effect of the modification framework on this relationship.

\begin{figure}[t]
  \centering
  \begin{minipage}[t]{0.3\linewidth}
    \vspace{0pt}
    \centering
    \includegraphics[scale=0.9]{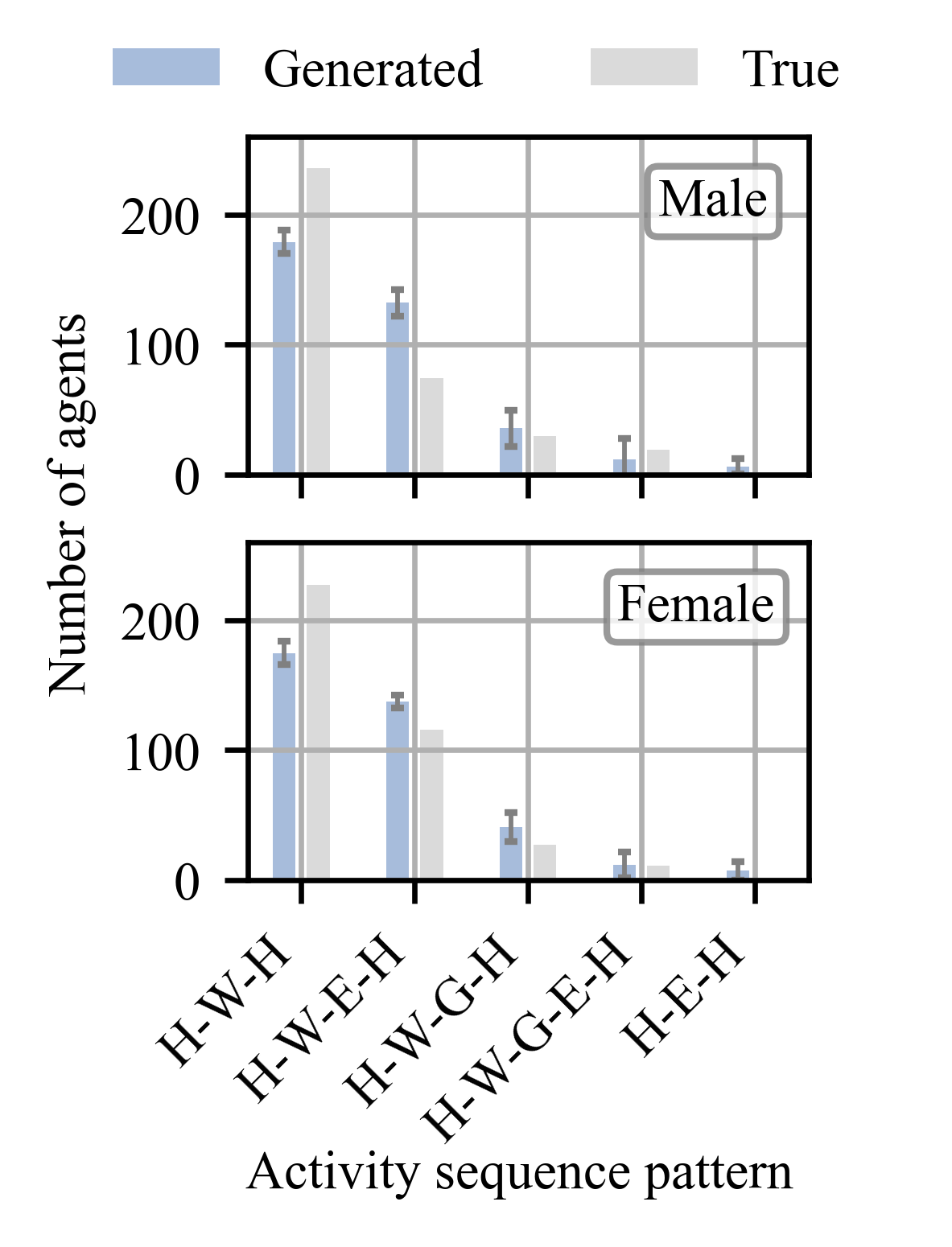}
  \end{minipage}
  \hfill
    \begin{minipage}[t]{0.69\linewidth}
      \vspace{10mm}
      \centering
      \footnotesize
      {\renewcommand{\arraystretch}{1.3}
      \begin{tabular}{l|rrr|rrr}
        \hline
        \multirow{2}{*}{\makecell{Activity \\ sequence $seq$}}
          & \multicolumn{3}{c|}{Male} & \multicolumn{3}{c}{Female} \\
        \cline{2-7}
          & $n^{seq}_{gen}$ & $n^{seq}_{true}$ & $e^{seq}_{prop}$ & $n^{seq}_{gen}$ & $n^{seq}_{true}$ & $e^{seq}_{prop}$ \\
        \hline
        H-W-H           & 179  & 236 & 0.159  & 175   & 227 &  0.136\\
        H-W-E-H         & 132  & 74  & 0.162  & 138   & 116 &  0.0577\\
        H-W-G-H         & 35.8 & 30  & 0.0162 & 41.0  & 27  &  0.0367\\
        H-W-G-E-H       & 11.6 & 19  & 0.0206 & 11.8  & 11  &  0.00210\\
        H-E-H           & 6.4  & 0   & 0.0178 & 7.2   & 1  &   0.0163\\
        \hline
      \end{tabular}
      }
    \end{minipage}
    
  \caption{Number of individuals by activity sequence pattern for full-time-employed males and females aged 18--75, after 10 iterations of activity location allocation and 10 iterations of activity modification. The generated data are shown as the mean over five runs, with bands indicating the minimum and maximum values. H means \texttt{\textbf{Home}}, W means \texttt{\textbf{Work}}, E means \texttt{\textbf{Entertainment}}, and G means \texttt{\textbf{Grocery}}.}

  \label{fig:activity_sequence_pattern}
\end{figure}

Figure~\ref{fig:activity_sequence_pattern} shows an example of the number of individuals by activity sequence pattern for full-time-employed males and females aged 18--75, after 10 iterations of activity location allocation and 10 iterations of activity modification.
We calculate the proportional error of activity sequence $seq$ as $e ^{seq} _{prop} = \frac{1} {n_{cat}} {|n^{seq}_{gen} - n^{seq}_{true}|}$ to assess distortion in activity sequence patterns, where $n_{cat}$ shows the total number of individuals in each category for male and female, $n^{seq}_{gen}$ and $n^{seq}_{true}$ show the number of individuals in each type of activity sequence $seq$ for the generated activity schedule and the true travel survey data, respectively.
The generated data show a distortion in activity sequence patterns with a proportional error of up to 16.2\% for males and 13.6\% for females.
This is expected because activity modification alters the activity sequences. 
Nevertheless, the distortion remains limited as the ranking of the most common activity sequences within each category is preserved.

\begin{figure}[t]
  \centering
  \begin{minipage}[t]{0.48\linewidth}
    \vspace{0pt}
    \centering
    \includegraphics[scale=0.9]{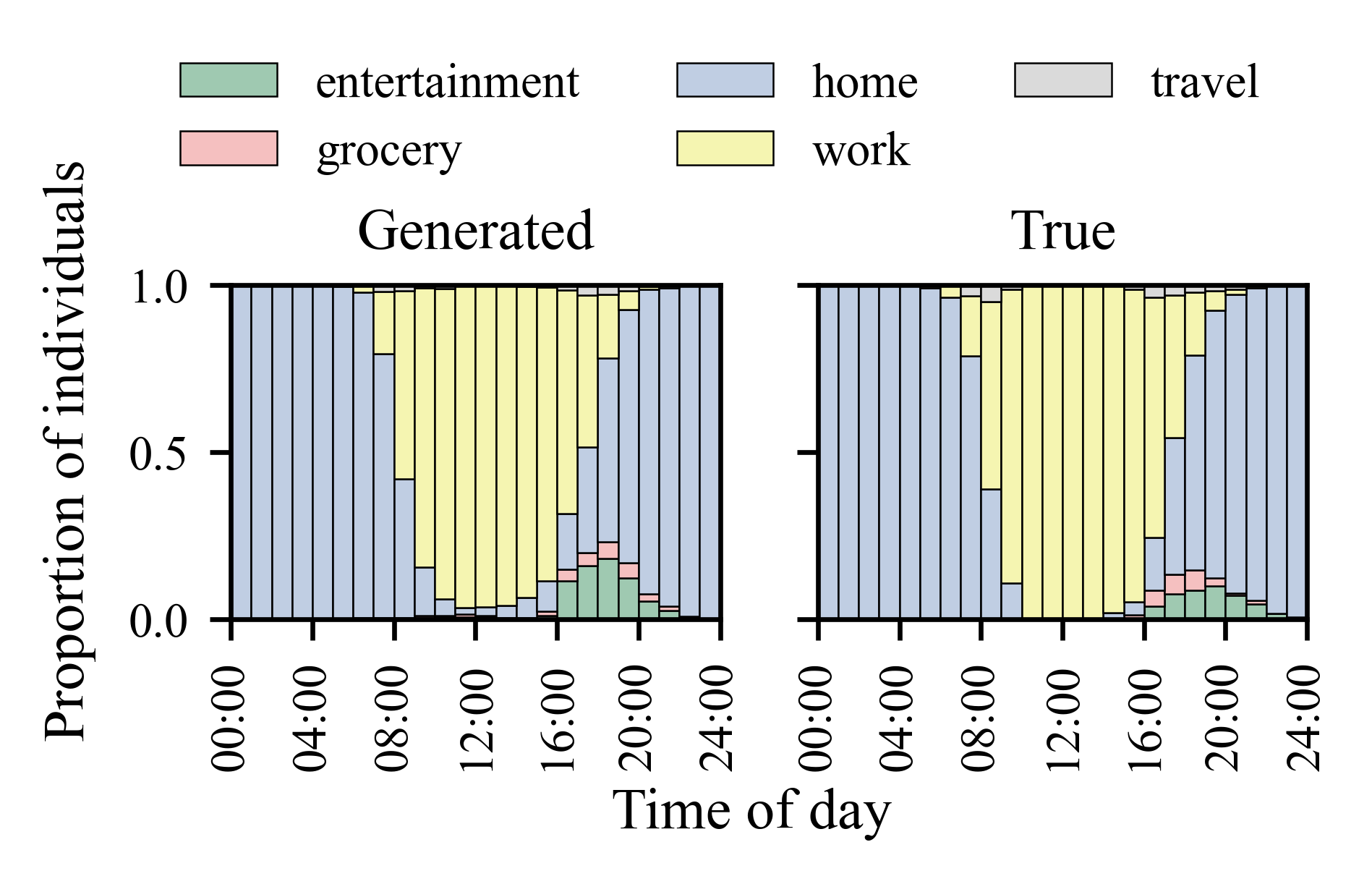}
  \end{minipage}
  \hfill
  \begin{minipage}[t]{0.48\linewidth}
    \vspace{7mm}
    \centering
    \begin{tabular}{lr}
      \hline
      Activity $act$& \makecell{$e ^{act} _{prop}$} \\
      \hline
      Home          & 0.032 \\
      Work          & 0.027 \\
      Grocery       & 0.0081 \\
      Entertainment & 0.015 \\
      \hline
      Overall average & 0.020 \\
      \hline
    \end{tabular}
  \end{minipage}

  \vspace{-5mm}

  \caption{Activity distribution across time slots for full-time-employed males aged 18--75, after 10 iterations of activity location allocation and 10 iterations of activity modification. The generated data are shown as the mean over five runs. The table reports the proportional error $e^{act}_{prop} = \sum_t \left| prop^{act,t}_{gen} - prop^{act,t}_{true} \right|$.}
  \vspace{-5mm}
  \label{fig:proportion_and_gap}
\end{figure}

Figure~\ref{fig:proportion_and_gap} presents an example of the activity distribution across time slots for full-time-employed males aged 18--75.
We calculate the proportional error as $e ^{act} _{prop} = \sum_t |prop ^{act,t}_{gen} - prop^{act,t}_{true}|$ to assess distortion in the activity distribution, where $prop ^{act,t}_{gen}$ and $prop^{act,t}_{true}$ show the proportion of individuals who engage in activity $act$ at time of day $t$ in the generated activity schedule and the true travel survey data, respectively.
A comparison between the generated data and the travel survey data indicates that the temporal activity distribution is not substantially altered, with a proportional error of 2.0\% on average.

\section{Discussion}
The results show that, through iterative refinement, the first part of the proposed simulated travel time based activity location allocation can reproduce realistic activity schedules while improving the matching between survey travel times and those generated by the traffic simulation by 46.7\% relative to the first iteration and incurs no distortion of activity sequence pattern.
A further, 5.5\% improvement is gained by applying a simplified version of an existing activity modification framework~\cite{GeoAI} -- limited to activity switching between individuals, rather than households, without considering the geospatial locations of switching candidates. However, the application of such framework incurs a proportional error in activity sequence patterns of up to 16.2\% in certain categories.
Nonetheless, the whole process preserves the characteristic activity sequences of individuals within specific sociodemographic groups, with an average change in proportion limited to 2.0\%.

As future work, the proposed method should be applied while considering households during activity switching. The representation of errors to flag activity modifications should be extended to indicate the geospatial areas where activity switching will lead to minimizing errors~\cite{GeoAI}.
Further, the proposed method should be validated using real world data. 
Previous studies have validated similar approaches using commuting origin--destination matrices, real world population data, and travel survey data, confirming the accuracy of activity schedule methods in real world case studies~\cite{HORL2021103291}.
Another important issue is computational cost. 
The location matching algorithm (Algorithm~\ref{alg:viterbi_route}) has a time complexity of $O(TV_n^2)$, where $T$ denotes the length of an individual's activity sequence and $V_n$ denotes the number of locations on the map. 
This computational burden may become problematic when the method is applied to larger maps or large scale simulations with many individuals. 
One possible solution is beam search, a heuristic method that prunes unlikely candidates during the search process. 
Because beam search is widely used to reduce the search space of the Viterbi algorithm and lower computational costs~\cite{ABDOU2004409}, incorporating it into Algorithm~\ref{alg:viterbi_route} may improve scalability.

%%
%% Bibliography
%%

%% Please use bibtex, 
\newpage
\bibliography{lipics-v2021-sample-article}

\end{document}